\documentstyle[aps,preprint]{revtex}
\begin{document}
\title{ Multi-Pion Correlation Effects on Two-Pion Interferometry
	\footnote{\it Partly supported by the National Science Foundation of
        China}}
\author{W.Q. Chao$^{a,b,c}$, C.S. Gao$^{a,c,d}$ and Q.H. Zhang$^{a}$}
\address{ a China Center of Advanced Science and Technology,
        (World Laboratory)
        P.O. Box 8730, Beijing 100080,  China\\
	b Institute of High Energy Physics, Academia Sinica, P.O. Box
	918(4),	Beijing 100039, China\\
	c Institute of Theoretical Physics, Academia Sinica, P.O. Box
	2735,	Beijing 100080, China\\
	d Physics Department, Peking University,
         Beijing 100871, China}
\vfill
\maketitle

\begin{abstract}
A general derivation of the multi-pion correlation function for completely
chaotic source is given. Its effects on
the pion multiplicity distribution, two-pion interferometry are studied.
A generalized multi-pion correlation function for a partially
coherent source is also discussed.
\end{abstract}

\section{  Introduction}

Hanbury-Brown and Twiss\cite{Hanb} were the first who applied the
Bose-Einstein correlations to measure the size of distant stars.  The
method was first applied to particle physics by Goldhaber et al.\cite{Gold}
in 1959.  Since then the size of the interaction region has been
measured by numerous experiments in high energy collisions using many
different types of particles\cite{Boal,zaj,Lor,Gyu}.
The basic idea of HBT (GGLP) effect is the following:  The intensity
 correlation function of the
two identical particles in momentum space
is related to the
Fourier transformation of the space-time distribution of the source.
Therefore using the two particle correlation function in momentum space,
one can extract the radius and lifetime of the source.
The source size measurement in high energy heavy-ion collisions has its
special meaning: Knowing the size of the source and the stored energy in
it one can obtain the corresponding energy density reached in the event,
which is one of the key quantities to investigate the possible
transition to quark-gluon
plasma (QGP) phase in high energy heavy-ion collisions.

In principle, the extension of such methods to more than two pions is
straightforward\cite{Gol2,Wills,zaj1,liu1,Biy1,wei1,zhang1,Cra1}.
Experimentally,
ultrarelativistic hadronic and
nuclear collisions provide the environment for creating dozens and in
some cases hundreds of pions\cite{Na35,Wa80,abb1,Na44}.
Therefore,  it is necessary to consider the multi-pion correlations in these
processes.  The Bosonic nature of pion should affect the single
and two pion spectrum and distort the two-pion correlation
function.  So it is very interesting to analyze the effects of multi-pion
correlation on these variables\cite{zaj1}.
Actually the bosonic nature also causes the
abundance of pion at small momentum.  Therefore, people are also interested in
analyzing the effects of multi-pion Bose-Einstein correlation on
the multiplicity distribution\cite{knox,pra1}.

In 1987, based on Monte-Carlo methods, Zajc analysed the effects of
multi-pion correlation on the phase-space and two-pion interferometry
\cite{zaj1}. Based on T matrix methods, Pratt analysed the effects of
multi-pion correlation on the pion multiplicity distribution\cite{pra1}.
In our paper, we give a more general derivation of the effects of
multi-pion correlation on two-pion interferometry.
The arrangement of this paper is as follows: The derivation of the multi-pion
correlation function for totally chaotic sources is given in section 2.
In section 3, the effects of multi-pion correlation on
the multiplicity distribution are discussed.
In section 4, the effects of multi-pion correlation on the two-pion
correlation function are analysed both for N pion events and for events with
all possible multiplicities.  In section 5, n pion correlation function for
a partially coherent source is derived. In section 6, as an example, our
results for a special source distribution are given, together with
some discussions.

\section{ N-pion correlation function for a chaotic source}

The general definition of the N-pion correlation
function $ C_{n}( \vec p_{1},\cdot\cdot\cdot \vec p_{n})$ is
\begin{equation}
C_{n}( \vec p_{1}, \cdot\cdot\cdot \vec p_{n})=
\frac{ P_{n}( \vec p_{1},\cdot\cdot\cdot, \vec p_{n})}
{\prod_{i=1,n} P_{1}(\vec p_{i})},
\end{equation}
where $ P_{n}(\vec p_{1}...\vec p_{n})$ is the probability of observing
$n$ pions with momenta$\{ \vec p_{i} \}$ all in the same event.
Here we derive a formula
for the correlation function which is valid for a totally chaotic source.

A state created by a classical pion source is described by\cite{Gyu,Chao}
\begin{equation}
|\phi>=exp({i\int d\vec p \int d^{4}x j(x) exp(ipx) c^{+}(\vec p)})|0>,
\end{equation}
where $c^{+}(\vec p)$ is the pion creation operator.  $j(x)$ is the current
of the pion, which can be expressed as
\begin{equation}
j(x)=\int d^{4}x' d\vec p A(x') j(x',\vec p) \nu(x') exp(ip(x-x')),
\end{equation}
here $j(x',\vec p)$ is the probability amplitude of finding a pion with
momentum $\vec p$,  emitted by the source at $x'$.
$A(x')$ is the probability amplitude of the source distribution.
$\nu(x')$ is
a random phase factor.  All sources
are uncorrelated in coordinate space when assuming:

\begin{equation}
<\nu^{*}(x')\nu(x)>=\delta^{4}(x'-x).
\end{equation}

The coherent state can expand in Fock-Space  as
\begin{equation}
|\phi>=\sum_{n=0,\infty}\frac{(i\int j(x)e^{ipx}c^{+}(p) d\vec p dx)^{n}}
{n!}|0>=\sum_{n=0,\infty}|n> ,
\end{equation}
with
\begin{equation}
|n>=\frac{(i\int j(x)e^{ipx}c^{+}(p) d\vec p dx)^{n}}{n!}|0> .
\end{equation}
Then we define the n-pion correlation function for n-pion state
\begin{equation}
C_{n}(\vec p_{1},.....\vec p_{n})=
\frac {<n|c^{+}(p_{1})\cdot\cdot\cdot c^{+}(p_{n})c(p_{n})
\cdot\cdot\cdot c(p_{1})|n>}
{\prod_{i=1,n}<1|c^{+}(p_{i})c(p_{i})|1>} .
\end{equation}

 From above definition,  the two-pion correlation function for two-pion
state can be written as

\begin{equation}
\begin{array}{lcl}
C( \vec p_{1}, \vec p_{2})&=& \frac{<2|c^{+}(\vec p_{1}) c^{+}(\vec p_{2})
c(\vec p_{2})c(\vec p_{1})|2>}{<1|c^{+}(\vec p_{1})c(\vec p_{1})|1>
<1|c^{+}(\vec p_{2})c(\vec p_{2})|1>}\\
	&&\\
  &=&1 + \frac{<1|c^{+}(\vec p_{1})c(\vec p_{2}|1>
<1|c^{+}(\vec p_{2})c(\vec p_{1})|1>}
{<1|c^{+}(\vec p_{1})c(\vec p_{1})|1>
<1|c^{+}(\vec p_{2})c(\vec p_{2})|1>}\\
\end{array}
\end{equation}

Noticing that
\begin{equation}
c(\vec p)|n>=i\int d^{4}x j(x) exp(ipx) |n-1>,
\end{equation}
using eq.(3),  we have
\begin{equation}
\begin{array}{lcl}
<1|c^{+}(\vec p_{1})c(\vec p_{2})|1>&=&
\int d^{4}x_{1} d^{4}x_{2} j^{*}(x_{1})j(x_{2})
exp(-i(p_{1}x_{1}-p_{2}x_{2}))\\
        &&\\
	&=&\int d^{4}x_{1}'d^{4}x_{2}' j^{*}(x_{1}',\vec p_{1})
	j(x_{2}',\vec p_{2})\nu^{*}(x_{1}')\nu(x_{2}')\\
	&&\\
	&&A^{*}(x_{1}')A(x_{2}')exp(-ip_{1}x_{1}'+ip_{2}x_{2}').\\
\end{array}
\end{equation}
Taking the phase average, we get
\begin{equation}
<1|c^{+}(\vec p_{1})c(\vec p_{2})|1>=
\int d^{4}x_{1}' j^{*}(x_{1}',p_{1})j(x_{1}',p_{2})|A(x_{1}')|^{2}
	exp(-i(p_{1}-p_{2})x_{1}').
\end{equation}
We define the semiclassical function
\begin{equation}
\begin{array}{lcl}
g(x',x,k)&=&\int d^{4}q j^{*}(x',k+q/2)j(x',k-q/2) exp(iq(x-x')).\\
\end{array}
\end{equation}
Here $k, q$ are four dimensional momenta.
$g(x',x,k)$ can be explained as the probability of finding a pion
with momentum $k$ at point $x$ due to the source at point $x'$.
Then we have
\begin{equation}
j^{*}(x',k+q/2)j(x',k-q/2)=\int d^{4}x g(x',x,k) exp(-iq(x-x')),
\end{equation}
and eq.(10) becomes
\begin{equation}
<1|c^{+}(\vec p_{1})c(\vec p_{2})|1>=\int d^{4}x g(x,k) exp(-iqx)
\end{equation}
Here $k=\frac{1}{2}(p_{1}+p_{2})$ , $q=p_{1}-p_{2}$ and
$g(x,k)$ is the probability of finding a pion  with momentum $k$
at point $x$.  It is defined as
\begin{equation}
g(x,k)=\int d^{4}x'\rho(x')g(x',x,k),
\end{equation}
where
\begin{equation}
\rho(x)=A^{*}(x)A(x)
\end{equation}
is the source distribution probability.
Inserting eq. (11) into eq. (5) gives the correlation function
\begin{equation}
C(\vec p_{1}, \vec p_{2})=1 + \frac{\int d^{4}x d^{4}x'g(x,k)g(x',k)
exp(iq(x-x'))}
{\int d^{4}x d^{4}x' g(x,p_{1}) g(x',p_{2})},
\end{equation}
where $q=p_{1}-p_{2}$, $k=(p_{1}+p_{2})/2$.

Similarly, the $n$-pion correlation function can be expressed as
\begin{equation}
C_{n}(p_{1},p_{2},\cdot \cdot \cdot p_{n})=\sum_{\sigma} \chi_{1,\sigma(1)}
\chi_{2,\sigma(2)}...\chi_{n,\sigma(n)}
\end{equation}
with the expression of $\chi_{i,j}$ as
\begin{equation}
\chi_{i,j}=\chi(p_{i},p_{j})=\frac{\int d^{4}x g(x, \frac{(p_{i}+p_{j})}{2})
e^{i(p_{i}-p_{j})x}}{\sqrt{\int d^{4}x d^{4}y(g(x,p_{i})g(y,p_{j})}}.
\end{equation}

In the following, we often use the expression of the n-pion
momentum probability distribution $S_{n}$ which can
be expressed as
\begin{equation}
S_{n}=\sum_{\sigma} \rho_{1,\sigma(1)}\rho_{2,\sigma(2)}...\rho_{n,\sigma(n)},
\end{equation}
with
\begin{equation}
\rho_{i,j}=\rho(p_{i},p_{j})=\frac{1}{n_{0}}
\int d^{4}x g(x, \frac{(p_{i}+p_{j})}{2})
e^{i(p_{i}-p_{j})x}.
\end{equation}

Here $n_{0}$ is the mean multiplicity without B-E correlation, defined as
\begin{equation}
n_{0}=\int g(x,k)d^{4}x d^{4}k.
\end{equation}
$\sigma(i)$ denotes the $i$th element of a permutation of the sequence
${1,2,3,\cdot \cdot \cdot, n}$, and the sum over $\sigma$ denotes the
sum over all $n!$ permutations of this sequence.

Once the source distribution g(x,k) is known, the multi-pion correlation
function can be calculated from eq.(18).  Therefore, $C_{n}$ can also
expressed by $S_{n}$ as
\begin{equation}
C_{n}(p_{1},p_{2},\cdot \cdot \cdot, p_{n})=\frac{S_{n}(p_{1},p_{2},
\cdot \cdot \cdot,p_{n})}{\prod_{i}\rho_{i,i}}
\end{equation}

\section{ Effects of multi-pion correlation on pion multiplicity
distribution}

 From the above definition of $|n>$, we have
\begin{equation}
\begin{array}{lcl}
\omega(n)=<n|n>&=&\frac{1}{n!}<0|\int j^{*}(x_{1})e^{-ip_{1}x_{1}}dx_{1}
\int j(x_{2})e^{ip_{1}x_{2}}dx_{2}dp_{1}\\
&&\\
&&\cdot \cdot \cdot
\int j^{*}(x_{2n-1})e^{-ip_{n}x_{2n-1}}dx_{2n-1}
\int j(x_{2n})e^{ip_{n}x_{2n}}dx_{2n}dp_{n}|0>,\\
\end{array}
\end{equation}
Following a similar derivation as in section two, we get
\begin{equation}
\omega(n)=<n|n>=\frac{n_{0}^{n}}{n!}\int S_{n}\prod_{i=1,n} d\vec p_{i}.
\end{equation}
Then we have
\begin{equation}
<\phi|\phi>=\sum_{k} <k|k>=\sum_{k}\frac{n_{0}^{k}}{k!}\int d\vec p_{1} \cdot
\cdot
\cdot d\vec p_{k} S_{k}.
\end{equation}
After considering the effect of multi-pion Bose-Einstein correlation, the
multiplicity distribution of pion becomes
\begin{equation}
P(n)=\frac{<n|n>}{<\phi|\phi>}=\frac{\omega(n)}{\sum_{k} \omega(k)}.
\end{equation}
For particles without BE Correlation, we have
\begin{equation}
\rho_{i,j}=\delta_{i,j}\rho_{i,i}
\end{equation}
therefore,
\begin{equation}
\int S_{n} \prod_{i} d\vec{p}_{i}=1
\end{equation}
and $P(n)$ reduces to the usual Poisson distribution.
 From above equation, we can also get the mean multiplicity $<M>$
\begin{equation}
<M>=\sum_{n}n\cdot P(n)
\end{equation}
\section{Effects of multi-pion correlation on
two-pion correlation function}
\subsection{ Two-pion correlation function for n-pion events}

For $n \pi$ events, the two-pion correlation function can be defined as
\begin{equation}
C_{2}^{n}=\frac{P_{2}^{n}(\vec p_{1},\vec P_{2})}{P_{1}^{n}
(\vec p_{1})P_{1}^{n}(\vec p_{2})} ,
\end{equation}
where $P_{2}^{n}(\vec p_{1},\vec p_{2})$ is the modified two-pion inclusive
distribution in $n$ pion events.  The
definition of $P_{i}^{n}(\vec p_{1},\cdot
\cdot \cdot,\vec p_{i})$ is
\begin{equation}
P_{i}^{n}(\vec p_{1},\cdot \cdot \cdot \vec p_{i})=
\frac{<n|c^{+}(\vec p_{i})\cdot \cdot \cdot
c^{+}(\vec p_{1})
c(\vec p_{1})\cdot \cdot \cdot c(\vec p_{i})|n>}{<n|n> n\cdot \cdot \cdot
(n-i+1)}.
\end{equation}

For an event with multiplicity
of $n$, the two-pion inclusive and single-pion inclusive distribution can be
expressed as
\begin{equation}
P_{2}^{n}(\vec p_{1},\vec p_{2})=\frac{\int \prod_{i=3,n} d\vec p_{i}
S_{n}(\vec p_{1}....\vec p_{n})}{\int \prod_{i=1,n} d\vec p_{i}
S_{n}(\vec p_{1}....\vec p_{n})},
\end{equation}
and
\begin{equation}
P_{1}^{n}(\vec p_{1})=\frac{\int \prod_{i=2,n} d\vec p_{i}
S_{n}(\vec p_{1}....\vec p_{n})}{\int \prod_{i=1,n} d\vec p_{i}
S_{n}(\vec p_{1}....\vec p_{n})}.
\end{equation}

As $n$ increases, the calculation of the intergration given above becomes
more and more complex. For the two-pion inclusive distribution in
the $n$ pion events,
there are only two-kinds of terms that we are
interested in.  One is
\begin{equation}
\cdot \cdot \cdot
\rho(p_{1},p_{k})\rho(p_{k},p_{l})\cdot \cdot \cdot\rho(p_{n1},p_{n2})
\rho(p_{n2},p_{2}) \rho(p_{2},p_{m})\rho(p_{m},p_{n3})\cdot \cdot \cdot
\rho(p_{n4},p_{i})\rho(p_{i},p_{1})\cdot \cdot \cdot,
\end{equation}
the other is
\begin{equation}
\cdot \cdot \cdot
\rho(p_{1},p_{k})\rho(p_{k},p_{l})\cdot \cdot \cdot \rho(p_{j},p_{1})
\rho(p_{2},p_{l})\rho(p_{l},p_{i1})
\cdot \cdot \cdot \rho(p_{m},p_{2})
\cdot \cdot \cdot.
\end{equation}

Now we define the function
\begin{equation}
G_{i}(p,q)=n_{0}^{i} \int \rho(p,p_{1}) d\vec p_{1} \rho(p_{1},p_{2})
d \vec p_{2} \cdot \cdot \cdot \rho(p_{i-2},p_{i-1})d \vec p_{i-1}
\rho(p_{i-1},q).
\end{equation}
The intergration in above term can be expressed as
\begin{equation}
\begin{array}{lcl}
\int \rho(p,p_{1})d\vec p_{1} \rho(p_{1},p_{2}) d\vec p_{2}
\cdot \cdot \cdot d\vec p_{k-1} \rho(p_{k-1},q)&&\\
       &&\\
\rho(q,p_{k+1})d\vec p_{k+1} \rho(p_{k+1},p_{k+2})d\vec p_{k+2}
\cdot \cdot\cdot d\vec p_{k+i-1}\rho(p_{k+i-1},p)&&\\
       &&\\
	=G_{k}(p,q)\cdot G_{i}(q,p)&&\\
\end{array}
\end{equation}
and
\begin{equation}
\begin{array}{lcl}
\int \rho(p,p_{1})d\vec p_{1} \rho(p_{1},p_{2}) d\vec p_{2}
\cdot \cdot \cdot d\vec p_{i-1} \rho(p_{i-1},p)&&\\
&&\\
\rho(q,p_{j+1})d\vec p_{j+1} \rho(p_{j+1},p_{j+2})d\vec p_{j+2}
\cdot \cdot \cdot d\vec p_{j+m-1}\rho(p_{j+m-1},q)&&\\
&&\\
=G_{i}(p,p)G_{m}(q,q).&&\\
\end{array}
\end{equation}

 From the expression of $S_{n}$
\begin{equation}
S_{n}=\sum_{\sigma} \rho_{1,\sigma(1)}\rho_{2,\sigma(2)}...\rho_{n,\sigma(n)}
\end{equation}
the number of terms of the form

$$\rho(p_{1},p_{k})\rho(p_{k}, p_{j})\rho(p_{j},p_{l})
 \cdot\cdot\cdot\rho(p_{m},p_{1})
\rho(p_{2},p_{i1})\rho(p_{i1},p_{im})\rho(p_{im},p_{in})
\cdot\cdot\cdot\rho(p_{ik},p_{2})C_{n-i}$$
is
\begin{equation}
\frac{(n-2)!}{(n-i)!}.
\end{equation}
$i$ is the number of $\rho$ that has not been included in $C_{n-i}$.
Similarly the number of terms of the form

$$\rho(p_{1}-p_{k})\rho(p_{k}-p_{l})
\cdot \cdot \cdot \rho(p_{j}-p_{2})\rho(p_{2}-p_{m})\cdot \cdot \cdot
\rho(p_{i1}-p_{1})C_{n-i}$$
is also
\begin{equation}
\frac{(n-2)!}{(n-i)!}.
\end{equation}

Then the two-pion inclusive distribution can be expressed as
\begin{equation}
P_{2}^{n}(\vec p,\vec q)=\frac{1}{n(n-1)}\frac{1}{\omega(n)}
\sum_{i=2,n}[\sum_{m=1,i-1}G_{m}(p,p)G_{i-m}(q,q)+G_{m}(p,q)
\cdot G_{i-m}(q,p)]\omega(n-i),
\end{equation}
with
\begin{equation}
\omega(n)=\frac{n_{0}^{n}}{n!}\int \prod_{k=1,n} d\vec p_{k} S_{n}.
\end{equation}

For the single-pion distribution,
we are only interested in the terms of the form
$\rho(p_{1},p_{k})\rho(p_{k},p_{j})\cdot\cdot\cdot\rho(p_{l},p_{1})C_{n-i}$
in the integrand, where $i$ is the number of $\rho$ that
are not included in $C_{n-i}$.
The number of this kind terms is
\begin{equation}
\frac{(n-1)!}{(n-i)!}.
\end{equation}
Then the single-pion distribution is
\begin{equation}
P_{1}^{n}(\vec p)=
\frac{1}{n}\frac{1}{\omega(n)}\sum_{i=1,n}G_{i}(p,p)\omega(n-i).
\end{equation}

Now the main question is to extract the expression of $\omega(n)$.
 From the expression of eq.(46), we have
\begin{equation}
\omega(n)=\frac{1}{n}\sum_{i=1,n}i\cdot C(i)\omega(n-i)
\end{equation}
with
\begin{equation}
C(i)=\frac{1}{i}\int d\vec p G_{i}(p,p).
\end{equation}

 From the above method the two-pion and
single pion inclusive distribution
can be calculated for $n$ pion events.

\subsection{ Two-pion correlation function for all events}

For a state $|\phi>$ which contains all possible multiplicities,
the modified single pion distribution can be expressed as
\begin{equation}
\begin{array}{lcl}
P_{1}^{\phi}(\vec{p})&=&\frac{<\phi|c^{+}(\vec p) c(\vec p)|\phi>}
{<\phi|\phi><M>}\\
&&\\
&=&\frac{\sum_{n}P_{1}^{n}(\vec p)\cdot n \cdot \omega(n)}{<\phi|\phi><M>}\\
&&\\
&=&\frac{\sum_{i}G_{i}(p,p)\sum_{n}\omega(n-i)}{<\phi|\phi><M>}=
\frac{\sum_{i}G_{i}(p,p)}{<M>},\\
\end{array}
\end{equation}
where eq.(26) is applied.  The two-pion inclusive
distribution can be expressed as
\begin{equation}
\begin{array}{lcl}
P_{2}^{\phi}(\vec{p},\vec{q})&=&
\frac{<\phi|c^{+}(\vec p)c^{+}(\vec q) c(\vec q)
 c(\vec p)|\phi>}
{<\phi|\phi><M(M-1)>}\\
&&\\
&=&\frac{\sum_{n}P_{2}^{n}(\vec p,\vec q) \cdot n \cdot (n-1) \omega(n)}
{<\phi|\phi><M(M-1)>}\\
&&\\
&=&\frac{\sum_{n}
[\sum_{i=2,n}(\sum_{m=1,i-1}G_{m}(p,p)G_{i-m}(q,q)+G_{m}(p,q)
\cdot G_{i-m}(q,p))\omega(n-i)]}{<\phi|\phi><M(M-1)>}\\
&&\\
&=&\frac{\sum_{i,j}G_{i}(p,p)G_{j}(q,q)+G_{i}(p,q)
\cdot G_{j}(q,p)}{<M(M-1)>}.
\end{array}
\end{equation}

The two-pion correlation function for all multiplicity distribution
is
\begin{equation}
C_{2}^{\phi}(\vec{p},\vec{q})=\frac{P_{2}^{\phi}(\vec{p},\vec{q})}
{P_{1}^{\phi}(\vec{q})P_{1}^{\phi}(\vec{p})}.
\end{equation}

$G_{i}(p,q), C(i)$ can be calculated either through Monte-Carlo
integration or by analytical integration.

\section{ n $\pi$ correlation function for a partially coherent source}

For a system with one coherent source and many other totally chaotic
sources, the state is described by
\begin{equation}
|\phi>_{part}=exp({i\int d\vec p \int d^{4}x (j(x)+
j_{c}(x)) exp(ipx) c^{+}(\vec p)})|0>,
\end{equation}
where $c^{+}(\vec p)$ is the pion creation operator.  $j_{c}(x)$ is the
current of the pion produced by coherent sources, while $j(x)$ is
the current produced by totally chaotic sources, which can be expressed as
eq.(3).
The state $|\phi>_{part}$ can be expanded as
\begin{equation}
|\phi>_{part}=\sum_{n}\frac{(\int (j_{c}(x)+j_{in}(x))e^{-ipx}c^{+}(p)dx)^{n}}
{n!}|0>=\sum|n>_{part} ,
\end{equation}
then the definition of the pion correlation function is unchanged.
For n-pion state, $|n>_{part}$, the n-pion correlation function can be
expressed as
\begin{equation}
C_{n}^{part}(p_{1},\cdot\cdot\cdot,p_{n})=
\frac{S_{n}^{part}}{\prod_{i=1,n}P_{1}^{part}(p_{i})},
\end{equation}
with
\begin{equation}
P_{1}^{part}(p)=\int g(x,p)dx +|j_{c}(p)|^{2} .
\end{equation}
The $S_{n}^{part}$ can be expressed as
\begin{equation}
S_{n}^{part}=S_{n}(\rho_{I})+\sum_{k=1,n}\frac{1}{k!}S_{n,k}(\rho_{I},
\rho_{c,k})
\end{equation}
with
\begin{equation}
S_{n,k}(\rho_{I},\rho_{c,k})=\sum_{l=1,\frac{n!}{k!(n-k)!}}
S_{n}^{l}(\rho_{I},\rho_{c,k}).
\end{equation}
Here $S_{n}(\rho_{I})$ is similarly defined as eq. (20) but with $\rho_{I}$
instead of $\rho$, the definition of $\rho_{I}$ is
\begin{equation}
\rho_{I}(p_{i},p_{j})=\int d^{4}x
g(x,\frac{p_{i}+p_{j}}{2})e^{i(p_{i}-p_{j})x}.
\end{equation}
$S_{n}^{l}(\rho_{I},\rho_{c,k})$ is similarly defined as
$S_{n}(\rho_{I})$ but with
$k$ terms $\rho_{c}$ instead of $\rho_{I}$.  Certainly, the total number of
this kind of substitutions is $\frac{n!}{k!(n-k)!}$.  $l$ represents one of
these substitutions.

The definition of $\rho_{c}(p_{i},p_{j})$ is
\begin{equation}
g_{c}(p_{i},p_{j})=j_{c}(p_{i})j_{c}^{*}(p_{j}).
\end{equation}

\section {Results and Discussions}

As an example, we choose the source distribution
\begin{equation}
g(x,p)=
\frac{1}{(\pi R_{0}^{2})^{3/2}}e^{-\frac{r^{2}}{R_{0}^{2}}}
\frac{1}{(2\pi m_{\pi}T)^{3/2}}e^{-\frac{E}{T}}\delta(t),
\end{equation}
and its non-relativistic form
\begin{equation}
g(x,p)=\frac{1}{(\pi R_{0}^{2})^{3/2}}e^{-\frac{r^{2}}{R_{0}^{2}}}
\frac{1}{(2\pi p_{0}^{2})^{3/2}}e^{-\frac{p^{2}}{2p_{0}^{2}}}\delta(t),
\end{equation}
where the parameter $R_{0}$ is the radius, T is the temperature and
$p_{0}^{2}=m_{\pi}T$.  $g(x,\frac{p+q}{2})$ can be approximately
expressed as
\begin{equation}
g(x,\frac{p+q}{2})=
\frac{1}{(\pi R_{0}^{2})^{3/2}}e^{-\frac{r^{2}}{R_{0}^{2}}}
\frac{1}{(2\pi m_{\pi}T)^{3/2}}e^{-\frac{E_{p}+E_{q}}{2T}}\delta(t)
=\frac{1}{(\pi R_{0}^{2})^{3/2}}e^{-\frac{r^{2}}{R_{0}^{2}}}
\frac{1}{(2\pi p_{0}^{2})^{3/2}}e^{-\frac{p^{2}+q^{2}}{4p_{0}^{2}}}
\delta(t) ,
\end{equation}
then we have
\begin{equation}
\rho(p,q)=\int g(x, \frac{p+q}{2})e^{i(p-q)x}dx
=\frac{1}{(2\pi p_{0}^{2})^{3/2}}
e^{-\frac{(p-q)^{2}R_{0}^{2}}{4}}e^{-\frac{p^{2}+q^{2}}{4p_{0}^{2}}}.
\end{equation}

Define
\begin{equation}
G_{n}(p,q)=\int \rho(p,p_{1})
\prod_{i=1,n-2} d\vec p_{i}
\rho(p_{i},p_{i+1})
d\vec p_{n-1}
\rho(p_{n-1},q)
\end{equation}
Using eq.(63),  we can easily get
\begin{equation}
G_{n}(p,q)=\alpha_{n} e^{-a_{n}(p^{2}+q^{2})+g_{n} \vec p \cdot \vec q}
\end{equation}
where
\begin{equation}
a_{n+1}=a_{n}-\frac{g_{n}^{2}}{4b_{n}}=\frac{R_{0}^{2}}{4}+\frac{1}
{4p_{0}^{2}}-\frac{R_{0}^{4}}{16b_{n}}
\end{equation}
\begin{equation}
b_{n}=a_{n}+\frac{1}{4p_{0}^{2}}+\frac{R_{0}^{2}}{4}
\end{equation}
\begin{equation}
g_{n+1}=\frac{g_{n}}{4b_{n}}
\end{equation}
and
\begin{equation}
\alpha_{n+1}=\alpha_{n}(\frac{1}{2p_{0}^{2}})^{3/2}(\frac{1}{b_{n}})^{3/2}
\end{equation}
with
\begin{equation}
a_{1}=\frac{R_{0}^{2}}{4}+\frac{1}{4p_{0}^{2}}, ~ g_{1}=R_{0}^{2}/2.
, ~ \alpha_{1}=\frac{1}{(2\pi p_{0}^{2})^{3/2}}.
\end{equation}
Then we have
\begin{equation}
C_{n}=\frac{1}{n}\int G_{n}(p,p)d \vec p=\alpha_{n}
\frac{\pi}{n(2a_{n}-g_{n})^{3/2}}.
\end{equation}

Now we can easily calculate the multiplicity,
single particle and two-particle
distribution of pions according to the formula given
in above sections.  In fig.1, we give the pion multiplicity distribution.
As the BE correlation is included, the probability with high pion
multiplicity is larger than the one of usual Poissonian
distribution. This feature is consistent with the nature of
Bosons.  We also find that as the radius and temperature decreases,
the effect of BE correlation on the multiplicity distribution
becomes larger.

The parameter $n_{0}$,  which is the mean multiplicity without
BE correlation, vs. the mean multiplicity $<M>$, is given in
fig.2.  When $M$ is not very large, $n_{0}$ increases with increasing
$<M>$.  However, for very large $<M>$, with increasing $<M>$
the corresponding $n_{0}$ is almost a constant.  In ref.\cite{pra1} to
obtain the Poissonian multiplicity distribution for a
system neglecting symmetrization, a factor $\frac{n_{0}^{n}}{n!}$ has to be
introduced,  while in our
work, starting from the classic multi-pion source the correction of
multi-pion BE correlation to multiplicity distribution is obtained and
it automatically reduces to a Poissonian when BE correlation is
neglected.

Including the multipion BE correlation, the single particle momentum
distribution for events with fixed multiplicity is given in fig.3.
It is shown that, as the multiplicity of pion increases,
the number of pions with low momentum becomes larger.
The effect of multi-pion correlation on the two-pion correlation
function are shown in
fig.4,  where we fixed the multiplicity of the events.
It can be seen clearly that as the multiplicity of the event increases,
the two-pion correlation function has a lower chaoticity, though
the actual source is totally chaotic.  The fitted radius also
becomes smaller when multiplicity increases as shown in fig.5.
This feature can be explained as follows: as the pion multiplicity
increases, the emission of pions is not
independent any more, multi-pion correlation is "Changed into" a factor
to affect the single-pion emission probability. This gives the source
some coherent property and makes the pion-source smaller.
It can be seen in fig.5 that the coherent parameter has
some correlations with the radius\cite{z2}. The lower coherent parameter
corresponds
to the smaller radius.  In fig.6, two-pion correlation function with
the mean multiplicity is shown.  A similar property as in fig.4 can be seen.
To obtain the correct behavior of the two-pion correlation function at high
mean multiplicities it is very important to have correct normalization
factors in eqs.(49-50), which ensure the correlation function
becoming flatter and its values approaching to one, when $<M>$ increases.

As shown in fig.2, the mean multiplicity increases very fast as $n_{0}$
approaches a certain value.  This seems to show that when $<M>$ becomes
very large, the "hardon" picture of pion may not be suitable any more,
and the system may undergo a kind of phase transition into the QGP phase.
However, the increase of $<M>$ with increasing $n_{0}$ should be very much
controlled by energy conservation which is not included in our formulation.
When energy conservation is included the behavior of $<M>$ at large $n_{0}$
may change dramatically.  Therefore, it is worthwhile to further
discuss the effects of multi-pion correlation on various observables
after including energy conservation\cite{Cha1}.

\begin{center}
{\bf Acknowledgement}
\end{center}

We wish to express our gratitude to Dr. Pang Yang for helpful discussion.

\newpage
\begin{center}
{\bf Figure Captions}
\end{center}

\end {document}